\documentclass{aa}  

\usepackage{graphicx}
\usepackage{adjustbox}
\usepackage{tabularx}
\usepackage{booktabs}
\usepackage{amsmath}
\usepackage{txfonts}
\usepackage{xcolor}
\usepackage{indentfirst}
\newcommand{\Msun}{\ensuremath{\, {M}_\odot}}

\begin{document}

   \title{The stellar to sub-stellar masses transition in 47 Tuc}

\author{C. Ventura\inst{1}, M. Tailo\inst{2,3}, P. Ventura\inst{1}, F. D’Antona\inst{1},
A. P. Milone\inst{2,4}, A. F. Marino\inst{2}, C. Fiumi\inst{5} }

   \institute{INAF, Observatory of Rome, Via Frascati 33, 00077 Monte Porzio Catone (RM), Italy \\ email: chiara.ventura@inaf.it \and
   Osservatorio Astronomico di Padova, Vicolo dell'Osservatorio 5, 35122 Padova, Italy \and
   Dipartimento di Fisica e Astronomia Augusto Righi, Università degli Studi di Bologna, Bologna, Italy \and
   Dipartimento di Fisica e Astronomia "Galileo Galilei", Univ. di Padova, Padova, Italy \and
   Dipartimento di Matematica e Fisica, Università degli Studi Roma Tre, via della Vasca Navale 84, 00100, Roma, Italy
   }


 
  \abstract
   {The study of the Globular Cluster 47 Tuc offers the opportunity to
   shed new light on the debated issue on the presence of multiple populations
   in Globular Clusters, as recent results from HST photometry and
   high-resolution spectroscopy outlined star-to-star differences in the
   surface chemical composition.}
   {The goal of the present investigation is the interpretation of recent
    JWST data of the low main sequence of 47 Tuc, in order to explore the stellar to
    sub-stellar transition, to derive the mass distribution of the individual 
    sources and to disentangle stars from different populations.}
   {Stellar evolution modelling of low-mass stars of metallicity $\rm [Fe/H]=-0.78$ 
   and oxygen content $\rm [O/Fe]=+0.4$ and $\rm [O/Fe]=0$ is used to
   simulate the evolution of the first and the second generation 
   of the cluster. The comparison between the calculated sequences with the data 
   points is used to characterize the individual objects, to split the different
   stellar components and to infer the current mass function of the cluster.}
   {The first generation of 47 Tuc harbours $\sim 45\%$ of the overall population
   of the cluster, the remaining $55\%$ making up the second generation. The
   transition from the stellar to the sub-stellar domain is found at 
   $\rm 0.074~M_{\odot}$ and $\rm 0.07~M_{\odot}$ for the first and second
   generations, respectively. The mass function of both the stellar generations are 
   consistent with a Kroupa-like profile down to $\rm \sim 0.22~M_\odot$.}
   {}

   \keywords{stars: evolution – stars: abundances – stars: interiors - globular clusters: individual: NGC 104
               }

   \titlerunning{The bottom of the main sequence of 47 Tuc}
   \authorrunning{C. Ventura et al.}
   \maketitle


\section{Introduction}

Observational evidence collected in the last decades demonstrated beyond any reasonable doubt that the majority of Galactic Globular Clusters (GC), if not all of them, harbour two or more stellar 
populations, differing in the distribution of the light elements abundance 
\citep{kraft94, gratton19, milone22} and, as deduced by the interpretation of the morphology of the horizontal branch (HB) and from the detection of multiple main sequences, of the helium mass fraction 
\citep{norris81, franca02, lee2005, dantona2005, piotto07, franca08, tailo20, franca22}.

These findings stimulated a lively debate regarding the action of a self-enrichment mechanism operating soon after the formation of GC, such that new stellar generations (second generation, hereinafter 2G) formed from the ashes of stars belonging to the first, original population (1G). Several candidates have been proposed as possible pollutants of the intra-cluster medium that might trigger the formation of multiple populations, namely fast rotating massive stars \citep{decressin}, massive binaries \citep{demink99}, 
super-massive stars \citep{pavel14} and massive AGBs \citep{paolo01}. 
The difficulties encountered by the various scenarios to account for the 
results of the observations were extensively discussed by \citet{renzini15}.

The investigations on this argument have traditionally focused on the
interpretation of the main sequence (MS) spread, the one observed 
in the red giant branch (RGB), and the morphology of the HB, in combination 
with the chemical patterns derived from high-resolution spectroscopy, 
to infer information on the presence of multiple populations in a
given GC, and to trace the history of the star formation during the 
infancy of the cluster. The most recent years have witnessed the development
of a new method, based on appropriate combinations of magnitudes
in different filters, which proves extremely powerful in disentangling
different stellar populations evolving in a given GC \citep{milone15, 
milone17, milone22}.

The advent of HST allowed us to extend the observations of GC stars
to the low MS, down to the brown dwarf domain
\citep{bedin01, richer02, richer06, richer08}. Interpretation of low-mass 
($\rm M < 0.2~M_{\odot}$) star data provides additional information 
regarding the presence of multiple populations in GC.
Indeed, the spectral energy distribution (SED) of these objects is extremely
sensitive to variations in element abundances, due to the relevant role
 molecular chemistry plays in the thermal stratification of 
low-temperature atmospheres \citep{marley02}. Furthermore, low-mass stars are 
fully convective and are exposed to minimal nuclear processing, given
the long time scales of hydrogen burning. The study of a brown dwarf is
equally interesting: unlike its higher mass counterparts ($\rm M > 0.07~M_{\odot}$) 
that reach the conditions for stable hydrogen burning, these 
objects undergo a gradual cooling process, evolving to lower effective temperatures 
and luminosities. The decrease in effective temperature favours the
formation of complex chemical compounds in the atmospheres, 
which eventually condense into liquid and solid states, with the
formation of clouds \citep{lunine86}. The extreme sensitivity of the 
spectra to chemical composition and age makes the analysis of the
brown dwarf a potentially efficient tool to discriminate among different
stellar populations and to infer the age of GC.

Interesting studies focused on the chemical diversity of GC stars in the 
$\rm M < 0.2~M_{\odot}$ mass domain were published, e.g., 
by \citet{dotter15,gera22a,gera22b}. Recent studies on the lower MS of M4 \citep{milone14} 
and NGC 6752 \citep{milone19} revealed that the relative distribution of stars belonging
to different stellar generations deduced from the analysis of the upper MS
and the RGB also characterises the lower MS. \citet{dondoglio} showed that
the fraction of stars in the different populations of NGC 2808
is constant throughout the mass range extending up to $\rm \sim 0.2~M_\odot$.

The innovative advent of JWST has further extended the possibility of investigating
the very low mass domain of GCs, rendering possible a detailed photometry of the 
stellar populations down to the brown dwarf domain. In this 
regard, \citet[][hereinafter MA24]{marino24}
have recently presented JWST data for 47 Tuc, covering the mass domain below 
$\rm \sim 0.1~M_\odot$. A thorough interpretation of the data set by MA24 
was hampered by the lack of $\rm M < 0.1~M_\odot$ stellar models, 
calculated for the chemical composition of the different stellar populations of 47 Tuc.

To overcome this difficulty, we calculated stellar models of mass 
$\rm M\geq 0.06~M_\odot$ of the same metallicity of 47 Tuc, based 
on two different oxygen content, chosen to span the $\sim 0.4$ dex spread 
in $\rm [O/Fe]$ indicated in MA24. We used these results
to characterise the sources observed in terms of mass and chemical composition.
Part of the effort is dedicated to reconstruct the present day mass function 
(MF) of the 1G and 2G of the cluster across the
star/sub-stellar domain discontinuity.

The paper is organised as follows: the description of ATON code for stellar
evolution used to calculate the evolutionary sequences, and the physical
and chemical ingredients adopted, is presented in section \ref{mod}.
The results obtained, with a discussion on the role of non-greyness and
on the separation of the sequences of stars with different chemistry in the
most relevant colour-magnitude planes considered, are presented in Sect.\,\ref{results}. In Sect.\,\ref{massfunction} we derive the mass function for the low main sequence and the transition masses, and in Sect.\,\ref{concl} we summarize and speculate on the results.

\section{Numerical, physical and chemical inputs}
\label{mod}
The stellar models used in the present work were calculated by means 
of the ATON code for stellar evolution, in the version described in \citet{paolo98}. 
Although the structures of the lowest mass stars and brown dwarfs are very simple, as they are fully convective, and therefore are close to polytropic structures, in principle, their study is actually hampered by a large number of uncertainties, explored in the years by more and more sophisticated approaches \citep[see, e.g.][, to quote only the historical paths.]{kumar1963, hayashi1963, dm1985, nelsonrapp1986, burrows1993, baraffe1995, montalban2000}. The uncertainties come from two different difficulties. 

First, below $\sim$4000\,K the atmospheric structure is increasingly dominated by complex molecules. Thus, the atmospheric structure ---and its temperature - density stratification--- requires including the complex physics of band formation. Calculation of atmospheric non--grey opacities is also complicated by the possible complex role of condensates \cite[e.g.][]{burgasser2002b}. The evolution with decreasing $\rm T_{eff}$ of the molecular band strength defines the classic M, L, T spectral types in the population I brown dwarfs \citep{kirkpatrick1999, burgasser2002a, burgasser2002b}, but it is still not fully understood in recent observations of the MS end of population II GCs \citep[starting from][for the lowest main sequence in the cluster M4]{dieball2016} and the associated model computation especially developed by \cite{gerasimov2020modelref}. 
The results for NGC\,6397 \citep{scalco63972024} and NGC\,6752 \citep{scalco67522024} are now complemented by those of 47\,Tuc \citep{scalco47tuc2025}. 
Difficult issues such as the role of condensates \citep{burgasser2002b, gera1} must be calibrated on the observations themselves. For the lowest masses, non ideal effects are important also in the atmospheric region. \\
\begin{figure}
\vskip-40pt
\begin{minipage}{0.48\textwidth}
\resizebox{1.\hsize}{!}{\includegraphics{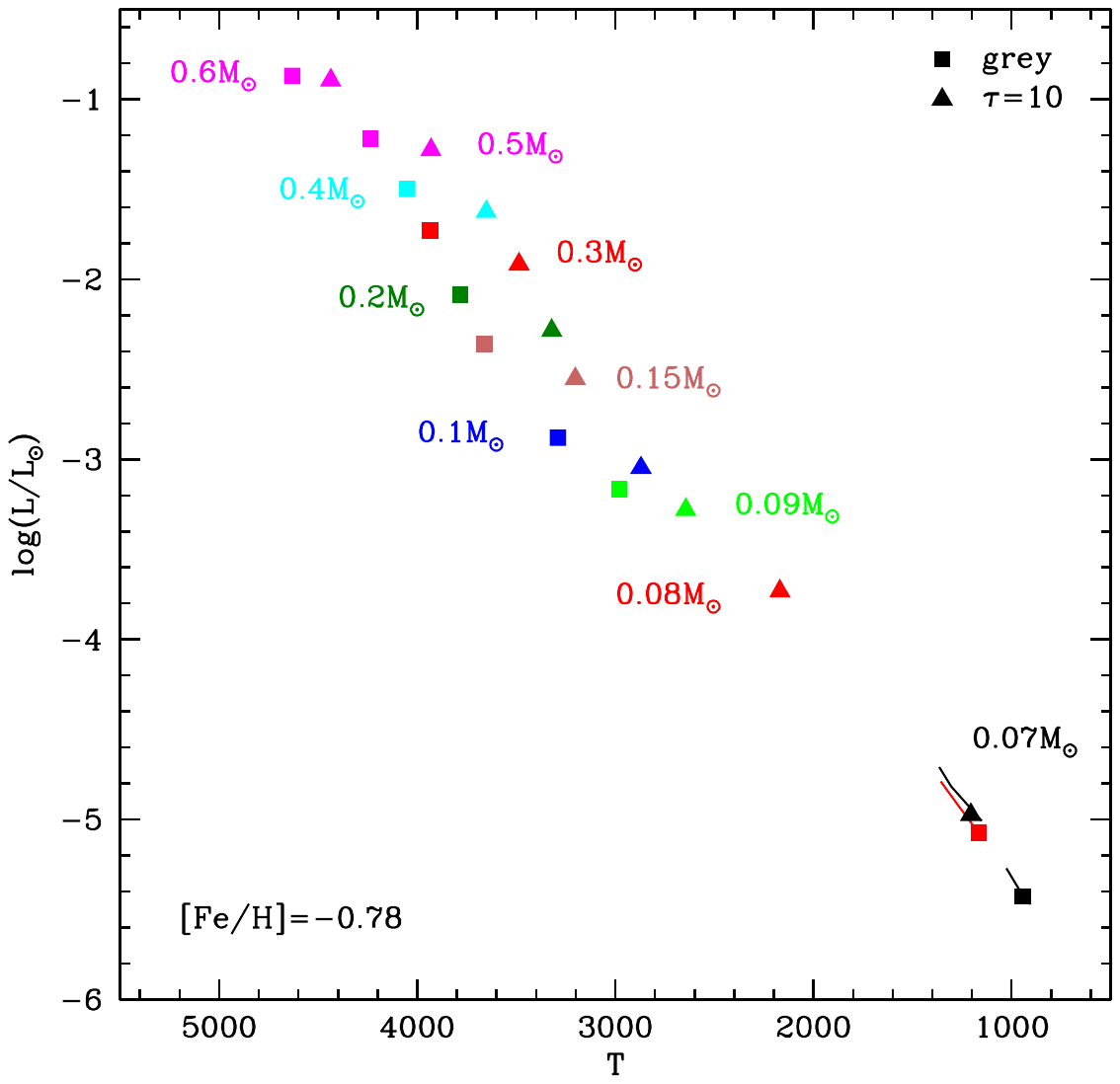}}
\end{minipage}
\vskip-60pt
\caption{Full points represent the position on the HR diagram of grey (squares) and non-grey (triangles) stellar models with mass in the $\rm 0.07 - 0.6~M_\odot$ range, at the age of $\rm 12~Gyr$. The colour highlights the same mass of the two sequences. For grey stellar models of mass $\rm 0.07~M_\odot$ and $\rm 0.08~M_\odot$, and for the non-grey $\rm 0.07~M_\odot$ model, the evolution in the age range from 8 to 12\,Gyr is displayed, the points indicating the position at $\rm 12~Gyr$.}
\label{fhr}
\end{figure}
\begin{figure}
\vskip-40pt
\centering
\includegraphics[width=0.49\textwidth]{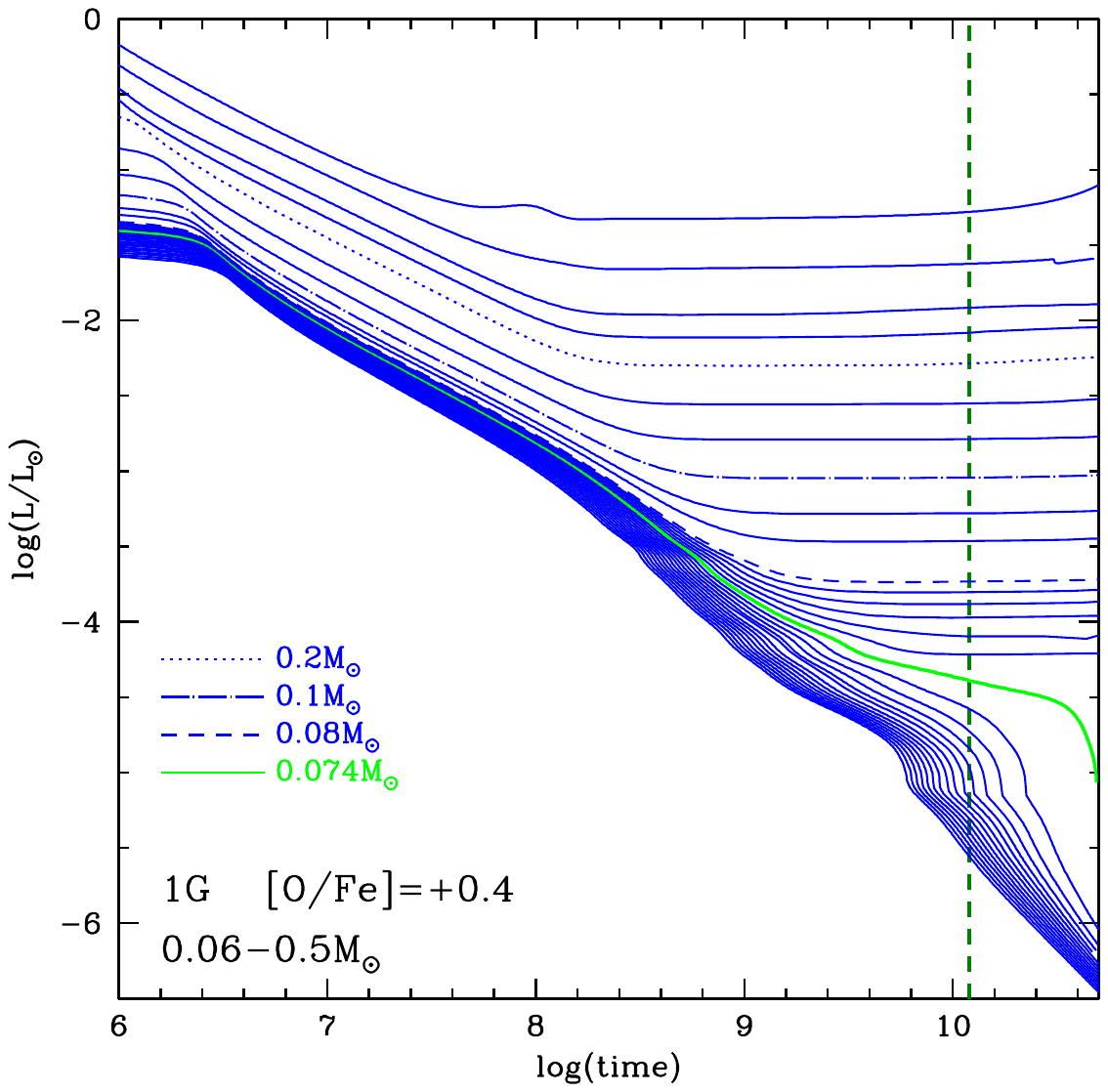}\hfil
\vskip-100pt
\includegraphics[width=0.49\textwidth]{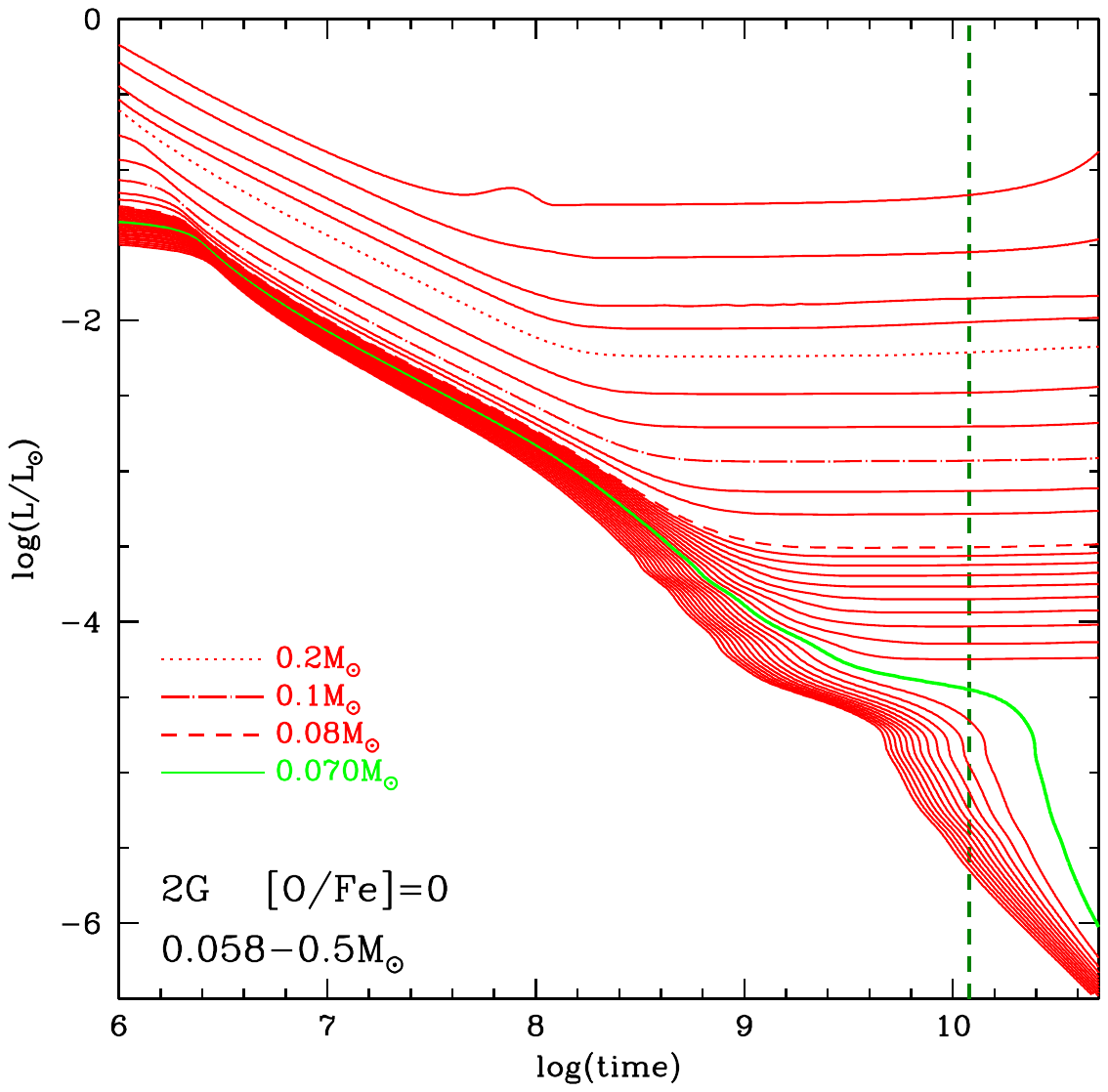}
\vskip-60pt
\caption{Time variation of the luminosity of stellar models calculated
with the physical and chemical input described in section \ref{mod}.
Stellar models representing the 1G and the 2G of the
cluster are shown  respectively in the top and bottom panel. The dashed 
green line indicates the age of $\rm 12~Gyr$. The tracks in green highlight 
the time evolution of the minimum mass supported by proton-proton chain
luminosity at the age of 12 Gyr, namely $\rm 0.074~M_\odot$ (top panel, 1G) and 
$\rm 0.070~M_\odot$ (bottom panel, 2G). The tracks of the stars of mass
$\rm 0.2~M_\odot$ (dotted lines), $\rm 0.1~M_\odot$ (dotted-dashed)
and $\rm 0.08~M_\odot$ (dashed) are highlighted.}
\label{ftimelum}
\end{figure}
\begin{figure}
\centering
\vskip-40pt
\includegraphics[width=0.49\textwidth]{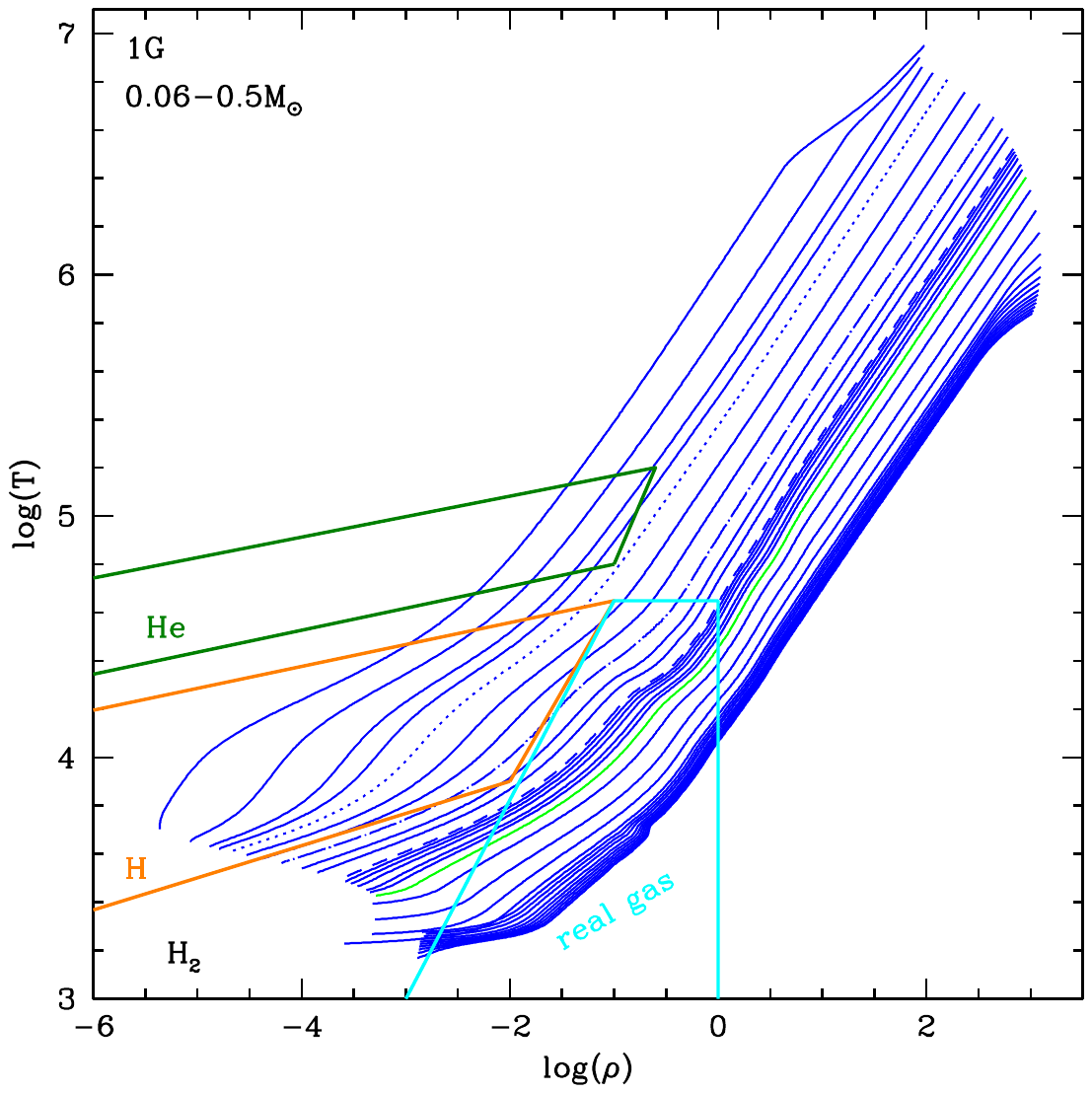}\hfil
\vskip-100pt
\includegraphics[width=0.49\textwidth]{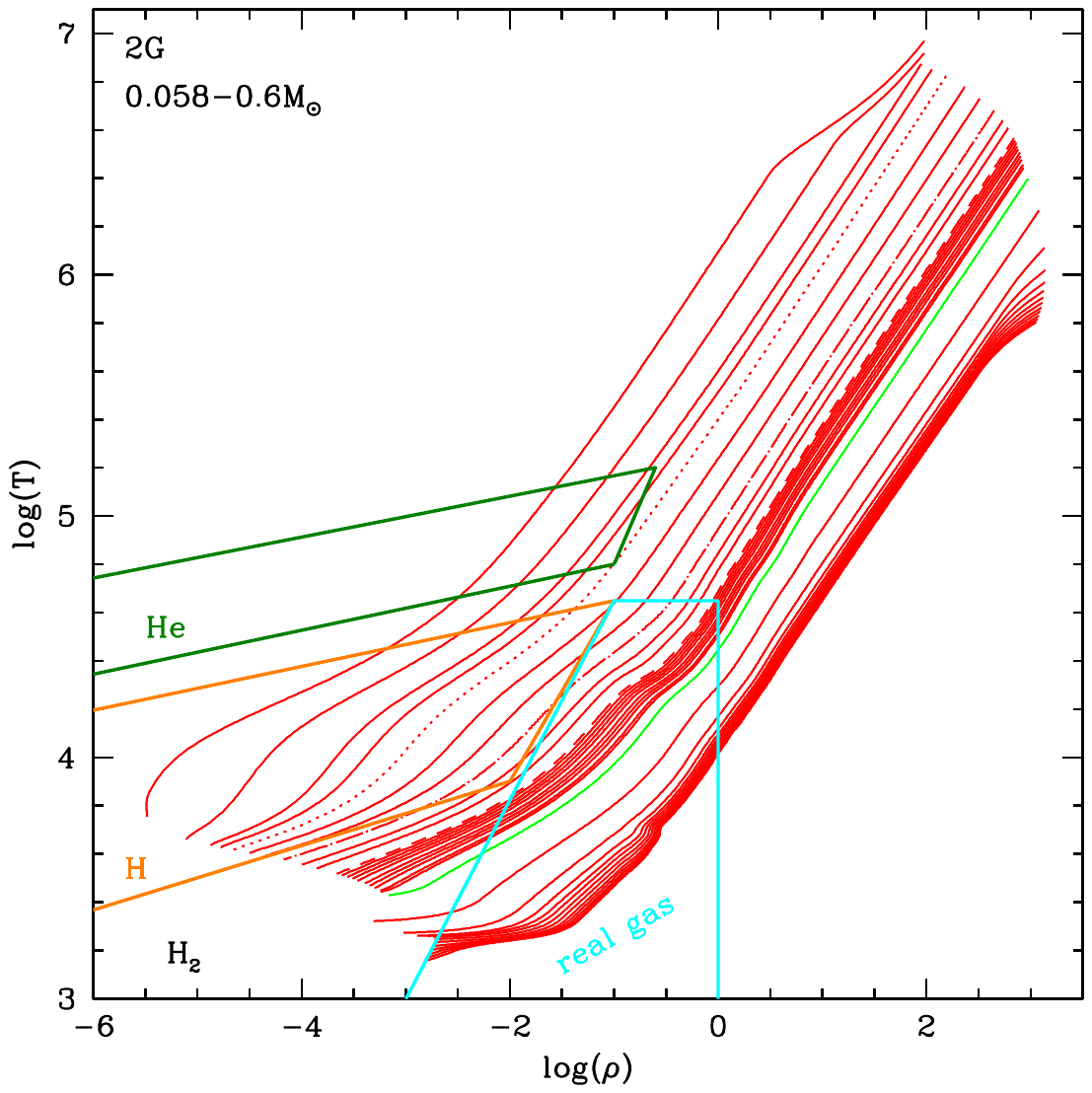}
\vskip-60pt
\caption{Thermodynamic structure of the stellar models of different
mass at the age of $\rm 12~Gyr$, in the density (g cm$^{-3}$) - temperature (K) plane.
The boundaries of the partial helium and partial hydrogen ionization regions are 
shown, as well as the region in which corrections to the ideal gas EOS become important, 
limited at high density by the pressure ionization boundary.
Top panel refers to 1G models, bottom panel refers to 2G. The same structures highlighted 
in Fig.2 are highlighted here. Below $\rm 0.08~M_\odot$ the mass step is $\rm 0.001~M_\odot$. This 
allows to see how fast in mass steps is the transition between stars and brown dwarfs.
}
\label{frhot}
\end{figure}
The second problem concerns the role of the equation of state (EOS) adopted to describe the interior. The most used EOS is at present the one developed by \cite[][SCVh EOS]{saumon1995}, for pure hydrogen, pure helium and pure carbon, for which the necessary intermediate compositions are obtained by the additive volume law, including considering carbon as an ``average'' metal. Only more recently a new EOS for pure H and pure He was calculated by \cite{chabrier2019}, and is still subject to the problem that it does not account for the interactions between the two species \citep[see, e.g. the discussion in][]{montalban2000}.

We tackle the first problem by adopting as boundary conditions the 
atmosphere sets computed by \citet{gera1}. The reason for this choice
is twofold. First, the atmospheric grids by \citet{gera1} are calculated
for the values $\rm [\alpha /Fe]=0, +0.4$, which bracket the range
of $\alpha-$enhancements deduced from the observations of 47 Tuc stars
\citep{marino16}: this allows for a thorough analysis in width of the cluster MS 
across the various colour-magnitude diagrams considered. 
Furthermore, use of boundary conditions by \citet{gera1} allows a 
straightforward comparison between the results obtained in the present 
work and those by \citet{scalco47tuc2025}, which are also based on 
the atmospheric models by \citet{gera1}.

For the EOS, we use the Tables of the ATON code \citep{paolo98} where the real gas regime is computed also for H-He mixtures, by matching the EOS by \cite[][(OPAL)]{iglesiasrogers1996} --including its updates until 2007-- with tables at high density computed following \cite{stolzmann1996, stolzmann2000}. 
A full description of the EOS tables is given in \cite{ventura2008}. 

\begin{figure}
\centering
\vskip -30pt
        \includegraphics[width=0.50\textwidth]{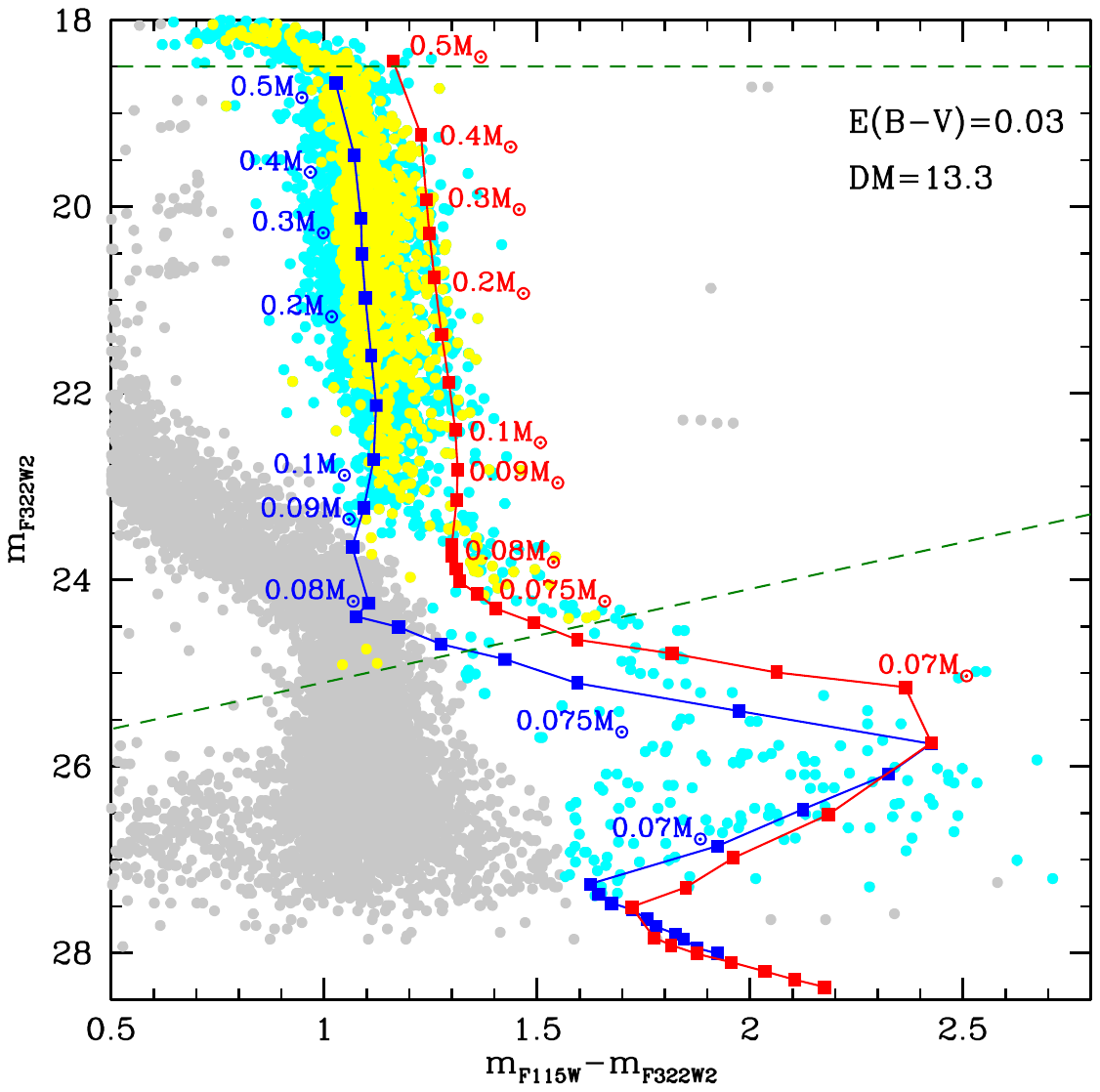}\hfil
\vskip-60pt
\caption{$\rm m_{F322W2}$\ versus $\rm (m_{F115W}-m_{F322W2})$\ 
colour-magnitude diagram of 47 Tuc stars from \citet{milone23} and MA24. Cyan dots indicate all sources observed for 47 Tuc, yellow dots indicate confirmed cluster members, according to the criteria described in section \ref{2gen} and grey dots indicate SMC members. The blue and the red lines indicate $\rm 12~Gyr$ isochrones for the 1G and 2G of the cluster, respectively. Some specific masses are labelled along the isochrones. 
The masses $\rm 0.5, 0.4, 0.3, 0.25, 0.2, 0.15, 0.12, 0.1, 0.09, 0.085, 0.08~M_\odot$, are shown. Below $\rm 0.08~M_\odot$ the models are spaced by $\rm 0.001~M_\odot$\ down to $\rm 0.06~M_\odot$ for the 1G and to $\rm 0.058~M_\odot$ for the 2G. 
Finally, the upper dashed green line represents the saturation limit ($\rm m_{F322W2}\simeq18.5$), while the lower diagonal green line indicates the limit where proper motions are available ($\rm m_{F115W}=26.1$ as in MA24, properly converted in $\rm m_{F322W2}$).
}
\label{fiso}
\end{figure}

\begin{figure}
\centering
\vskip -30pt
\includegraphics[width=0.50\textwidth]{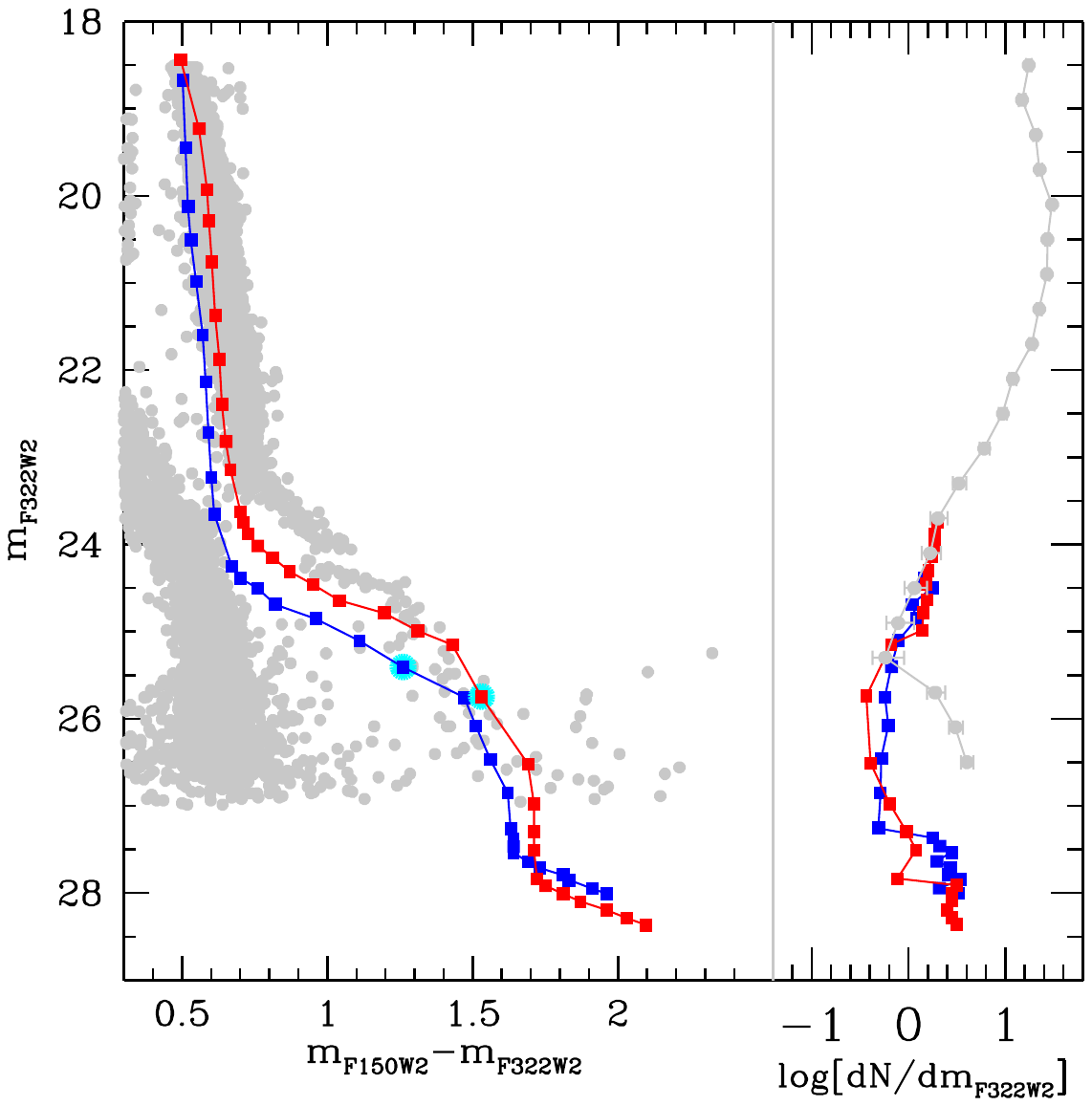}
\vskip-60pt
\caption{Colour magnitude data in the $\rm (m_{F150W2}-m_{F322W2}, m_{F322W2})$ plane (left,  grey dots), are compared with the theoretical models. Contrary to the previous diagram, the agreement is unsatisfactory. These are the data on which the luminosity function has been derived. The LF is shown in the right side of the figure (grey dots), together with the theoretical luminosity function obtained by assuming a flat mass function below $\rm 0.3~M_\odot$, by normalizing the mass luminosity derivative of the models to the data. The figure shows that the observational LF would be reproduced if the transition masses were more luminous that computed by about 1.5\,mag. 
}
\label{doppio}
\end{figure}

The evolutionary sequences were started from an early evolutionary
pre-MS phase, corresponding to the central temperature of 
$5 \times 10^5$ K, and were extended until the end of core 
hydrogen burning; for objects not reaching full support by nuclear energy generation,
the computations were stopped for age $<$50\,Gyr, or below the 
threshold covered by the atmospheric grids (at $\rm T_{eff}=500~K$). 
The range of masses is  in the $\rm 0.06-0.8~M_\odot$ 
domain\footnote{The choice of the starting point has no
effects on the conclusions reached, considering that in the low-mass
domain explored in the present work the duration of early evolutionary 
phases is negligible when compared to that
of the MS lifetime. Also, for stars in the sub-stellar mass
domain the starting point of the evolution is not relevant,
because the cooling time scales are of the order of a few Gyr,
while the initial contraction phase during which the central
temperature increased proceeds with time scales below
$10^8$ yr.}.

We computed two sets of models, taking advantage of the available 
atmospheric grids by \citet{gera1}. The first set assumes the standard chemical 
composition of the 1G of 47 Tuc, namely $\rm [Fe/H]=-0.78$, and alpha enhancement 
$\rm [\alpha/Fe]=+0.4$ \citep{marino16}. Based on the solar mixture by 
\citet{asplund09}, these values correspond to the metallicity $\rm Z=0.0047$. 
The initial helium is taken as $\rm Y=0.255$. A second set of tracks simulates the evolution of the stars belonging  to the 2G of 47 Tuc: in this case we assume a depletion in the oxygen content, so that $\rm [O/Fe]=0$ \citep{marino16}, and a helium mass fraction $\rm Y=0.30$, in agreement with the analysis of the spread of helium in 47 Tuc by \citet{milone18}. 

We considered different locations of the matching region between the interior and the exterior, located at optical depth $\tau=1, 3, 10, 100$, as described in detail in \citet{montalban2004}. 
We finally adopted $\tau$=10 as boundary for the interior computations, as a compromise to have a good description of the non grey atmosphere and the necessity of having an EOS including the correct description of the real gas for the smallest masses.\\
The reason why we limit our exploration to $\rm [O/Fe]=0$ and $\rm [O/Fe]=+0.4$ is because these are the oxygen abundances for which the model atmospheres by \citet{gera1} are available. This limitation prevents a more detailed study of the different stellar populations in 47\,Tuc, but it proves sufficient to undertake a broad distinction between 1G and 2G stars and to investigate the behaviour of models at the bottom of the main sequence.
For the sake of comparison, we also compute the evolutions also for a grey atmosphere, where the matching of the results from the solution of the equations of stellar equilibrium from the interior and those from the modelling of the atmosphere is done at the optical depth $\tau=2/3$. 

\section{Results}
\label{results}
The low main sequence structures, according to the arguments discussed in the previous section, demand the use of detailed non-grey atmospheric and sub-atmospheric modelling. Furthermore, when the HR diagram is moved to the observational planes, the effects of the absorption bands related to the presence of many specific molecular species, particularly water, affect the relation between the effective temperature of the star and the spectral energy distribution.

In this section we first discuss the differences between the results obtained with the grey and non-grey treatments of the external regions, then we focus on the possibility of disentangling  the multiple populations of 47 Tuc by interpreting the distribution of the stars in the colour-magnitude diagram.

\subsection{The role of non-greyness in the theoretical HR diagram}
Fig.~\ref{fhr} shows the position on the HR diagram at the age of
$\rm 12~Gyr$ of grey (squares) and non-grey (triangles) stellar models for masses in the 
$\rm 0.08-0.6~M_\odot$ range. For the lowest masses below
$\rm 0.08~M_\odot$ we show the evolutionary tracks covering the ages from
8 to 12 Gyr, with the location at $\rm 12~Gyr$ indicated. 
For clarity reasons, for the non-grey models we show the $\tau=10$ case only, 
as the results obtained with different $\tau$'s are extremely similar
(within 20~K) each other.

The MS luminosity decreases with the stellar mass for $\rm M \geq 0.08~M_\odot$:
H-burning sets at $\rm \sim 6.4\times10^{-2}~L_\odot$
for $\rm M = 0.5~M_\odot$, $\rm \sim 7.6\times10^{-3}~L_\odot$ 
for $\rm M = 0.2~M_\odot$, $\rm \sim 1.3\times10^{-3}~L_\odot$
for $\rm M = 0.1~M_\odot$. The lower mass domain 
is characterized by the growing influence of core electron degeneracy, 
which prevents efficient heating of the central regions following 
the increase in densities.
In the very low-mass domain the central temperatures are low enough that H-burning is fully inhibited,
and the ``brown dwarfs" cool along approximately constant radius
sequences with decreasing luminosities.
There is a very small range of masses in which p-p burning is ignited, 
but the nuclear luminosity is not able to fully support the stellar luminosity, 
and these ``transition objects" slowly decrease in luminosity on timescales of 
billion years \citep{dm1985}, see Fig.~\ref{ftimelum}.

The main effect of non-greyness in the theoretical HR diagram of Fig.~\ref{fhr} is that 
$\rm M > 0.1~M_\odot$ stellar models are cooler than the grey counterparts. 
This is particularly important for $\rm 0.1-0.2~M_\odot$ stars, 
where the differences $\rm \Delta T_{eff}$ in the effective temperature are 
slightly below  $\sim 500$ K. For higher masses, we find that 
$\rm \Delta T_{eff}$ is $ \sim 250$ K at $\rm 0.5~M_\odot$ (and nearly the same 
for $\rm 0.6~M_\odot$), then becomes negligible for higher masses. 
The differences are due to the fact that the Eddington 
approximation results in an atmospheric profile substantially
cooler and denser than the non-grey profile, so a hotter effective 
temperature  matches the internal adiabatic stratification, as 
extensively discussed in the literature, see e.g. \citet{chabrier97}.

The treatment of the boundary conditions also affects the threshold mass
separating the stars that reach the MS from those that follow a
sub-stellar behaviour, evolving along a constant radius track towards 
fainter luminosities. The threshold mass is $\rm \sim 0.08~M_\odot$ and 
$\rm \sim 0.073~M_\odot$ for the grey and non-grey cases, 
respectively. Therefore\footnote{Thus, in Fig.~\ref{fhr}, we show the evolution of the 
$\rm 0.08~M_\odot$ grey stellar model, while for the  non-grey counterpart of the same mass,  we show only the main sequence location at $\rm 12~Gyr$.}
The impact of non-greyness tends to vanish in the sub-stellar domain, as 
the evolution of these stars, which runs along constant radius sequence, is 
essentially determined by the stellar mass. Non grey model atmospheres anyway 
are the fundamental input to derive colours, and in the following we will deal 
only with the non-grey models, whose inputs are described in section \ref{mod}.

In Fig.\,1 the most interesting and well known feature is the gap in luminosity at the boundary between stars and transition masses, which catches the eye as a sudden increase in the luminosity difference between adjacent models for the same mass difference (compare the luminosity gap between the $\rm 0.08~M_\odot$ and $\rm 0.07~M_\odot$ non-grey models and the much smaller gap between the $\rm 0.08~M_\odot$ and $\rm 0.09~M_\odot$ model).

\subsection{Structure and evolution of 1G and 2G stars}
The temporal evolution in luminosity of 1G and 2G stellar models is shown in
Fig.~\ref{ftimelum}. The luminosity decreases during the initial 
evolutionary stages, when the stars are supported by gravitational contraction\footnote{Below $\rm 0.2~M_\odot$, the initial Deuterium burning phase lasts for 2\,Myr or longer, and appears in the figure as a temporary stop in luminosity during the contraction phase.}. 
The global behaviour is sensitive to the mass of the star, particularly
on whether the mass is above the threshold required to start H burning. 
The tracks reported in Fig.~\ref{ftimelum} show a clear separation
between the higher masses, in which stable hydrogen burning is activated,
and the luminosity remains constant, and the smaller masses, in which
the hydrogen burning power ($\rm L_{nuc}$) remains lower the stellar luminosity, so that 
it continues to decrease. 
The minimum mass reaching $\rm L_{nuc}=L_{tot}$ is $\rm 0.074~M_\odot$ for the 1G and 
$\rm 0.070~M_\odot$ for the 2G.  
This difference in mass can be attributed to the different initial helium mass fractions for the two sets of models. To verify this, we compared models with the same helium content Y=0.255 of 1G and 2G, and found that the atmospheric boundary conditions were not very different despite the very different [O/Fe] content of the two mixtures. We also computed 2G models for different Y values and report the colours and luminosities of the
minimum mass that reach the $\rm L_{nuc}=L_{tot}$ condition in Table \ref{tab:elio}. The minimum masses and luminosities decrease slightly with increasing helium content.

\begin{table*}[htb]
\caption{Transition mass and properties for different chemistries}
\centering
\setlength{\tabcolsep}{10pt}
\renewcommand{\arraystretch}{1.5}
\begin{tabular}{c c c c c c c}

\hline 
$\rm Y$ & $\rm M/M_\odot$ & $\rm T_{eff}$ & $\rm log(L/L_\odot)$ & $\rm m_{F322W2}$ & 
$\rm m_{F115W}-m_{F322W2}$ & $\rm m_{F150W2}-m_{F322W2}$  \\
\hline
$\rm 1G, Y=0.255$ & $0.074$ & $1581$ & $-4.39$ & $12.10$ & $1.95$ & $1.25$ \\
$\rm 2G, Y=0.28$  & $0.073$ & $1701$ & $-4.26$ & $11.75$ & $2.06$ & $1.31$ \\
$\rm 2G, Y=0.30$  & $0.071$ & $1712$ & $-4.29$ & $11.79$ & $2.11$ & $1.34$ \\
$\rm 2G, Y=0.35$  & $0.066$ & $1630$ & $-4.37$ & $11.89$ & $2.24$ & $1.4$  \\
\hline
\end{tabular}
\tablefoot{The table lists various theoretical and observational properties 
of stellar models of minimum mass that reach the $\rm L_{pp}=L_{nuc}$
at the age of 12 Gyr, for different chemical composition. Col. 1 gives the helium 
content and if the composition refers to 1G or 2G, col. 2 and 3 give $\rm T_{eff}$ and
luminosity, whereas col. 3-6 report $\rm M_{F322W2}$ and the $\rm (m_{F115W}-m_{F322W2})$ 
and $\rm (m_{F150W}-m_{F322W2})$ colours. 
}
\vskip10pt
\label{tab:elio}
\end{table*}

The luminosity differences between stellar models of the same mass 
belonging to the two generations decrease from $\sim 40\%$, 
for $\rm 0.3-0.5~M_\odot$ stars, to $\sim 20\%$, down to
$\rm 0.1~M_\odot$; the differences increase again in the low-mass 
domain, because the threshold masses for the ignition of hydrogen
differ, thus stars of the same mass behave differently when the
stellar$/$sub-stellar domain is approached.

The time evolution of the luminosity for the lowest mass range examined 
shows also a non linear behaviour dependent on the treatment of the equation 
of state into these high density -- low temperature structures. 
The interior structure of the stellar models of different mass taken at 
the age of $\rm 12~Gyr$ is shown in Fig.~\ref{frhot} for both the 1G and 2G.

The regions of molecular hydrogen and helium partial ionization are 
highlighted in the figures, together with the region in which the real gas effects 
become important in the structure. The latter region is limited by the lines of 
temperature and pressure ionization. The drops in the central temperature below 
$\rm M=0.073~M_\odot$ (top panel of Fig.~\ref{frhot}, 1G) and $\rm M=0.07~M_\odot$ 
(bottom panel, 2G)
characterize the transition between the stellar and substellar regime, and it 
appears clear in the figures that, in addition to the boundary conditions, also 
the EOS description in the real gas regime may affect the mass location of the transition.

Figures \ref{ftimelum}, and mostly Fig.~\ref{frhot} show in an interesting way the 
transition between stars and brown dwarf regimes. The masses plotted below 
$\rm M=0.08~M_\odot$ are spaced by $\rm 0.001~M_\odot$, and we see that in the log(T) 
versus log($\rho$) plane they are first very close together, then more distant, 
and finally very close again. The region in which they become more distant corresponds 
to the transition between stellar and substellar structures, which is covered by a very small range of masses $\rm \sim 5 \times 10^{-3}~M_\odot$.

\subsection{Disentangling the stellar populations in the low main sequence}
\label{2gen}
The difference in the oxygen content between the stars belonging to the 1G
and 2G of 47 Tuc, though not particularly relevant for the determination of the
position of the stars in the theoretical HR diagram, definitively affects the distribution 
in the observational colour magnitude diagrams, specifically in those using filters overlapping 
with the absorption features associated to the presence of oxygen-bearing molecules 
(primarily water). Such features are deeper in the spectra of 1G stars, given the higher 
oxygen mass fraction, than in the 2G counterparts. This effect arises at effective 
temperatures below $\rm 4000~K$, becoming more and more important as $\rm T_{eff}$
decreases \citep{gera1}.

Following \citet{milone23} and MA24, we focus on the 
$\rm (m_{F115W}-m_{F322W2}$, $\rm m_{F322W2})$ and
$\rm (m_{F150W}-m_{F322W2}$, $\rm m_{F322W2})$ colour-magnitude diagrams. 
These filters combination maximises the impact of the oxygen variation between 
the populations, as the filters cover the spectral region where the flux 
difference between 1G and 2G stars is at its maximum. 

The flux of 2G stars is higher than that of 1G stars in the spectral region covered
by the $\rm H_2O$ bands, the differences being of the order of $\rm 30\%$. 
The largest differences are found in the wavelength intervals 
$\rm \lambda \sim 2.6 - 2.8 ~\mu m$, including the median wavelength of $\rm F322W2$ filter.

Tables~\ref{tab:teflum1g} and \ref{tab:teflum2g} report the effective temperature and
luminosity of the model stars belonging to 1G and 2G of the cluster,
respectively, the $\rm (m_{F115W}-m_{F322W2})$ and
$\rm (m_{F150W}-m_{F322W2})$ colours, and $\rm m_{F322W2}$, at the age of 12 Gyr.
The last two columns in the tables give the fraction luminosity of the star due 
to proton-proton reactions at the age of 12 Gyr, (unity means that the star 
reaches stable core H-burning conditions), and the maximum
value attained by the same quantity during the evolution.




In Fig.~\ref{fiso} we show the distribution of 47 Tuc sources by MA24 in the 
$\rm (m_{F115W}-m_{F322W2}, m_{F322W2})$ diagram. 
Cluster members along the low MS, reported as yellow points, have been selected 
according to the indication coming from the proper motions, as reported in \citet{milone23} 
and MA24. In a nutshell, we select as cluster members all stars brighter than the 
$\rm m_{F115W}=26.1$ mark, for which proper motion have low uncertainty, 
and on the left of the $\rm DR=2.1$ mark. We also include in the CMD stars that 
do not have proper motions yet: these sources are indicated with cyan dots 
in Fig.~\ref{fiso}\footnote{We note that \citet{scalco47tuc2025} 
limit their analysis to $\rm M\geq0.1\Msun$, corresponding to $m_{\rm F322W2}\leq23\,mag$. 
This choice is related to the set of isochrones adopted, which unlike ours do not
extend down to beyond the stellar$/$non stellar domain, and to the availability 
of direct spectroscopic estimates of the chemical abundances of very low-mass stars.}.

Overlapped to the data points we report the $\rm 12~Gyr$ isochrones for the 1G and
the 2G population of the cluster: solid squares along each isochrone indicate
the position of stellar models of given mass. A satisfactory agreement
between the theoretical isochrones and the data points is obtained by assuming 
reddening $\rm E(B-V)=0.03$ and distance modulus $\rm (m-M)_0=13.3$: 
this choice allows to reproduce the excursion toward the red of the data points at 
$\rm m_{F322W2} > 24$.

Starting from the brighter region of the plane, the 1G and 2G sequences first 
move approximately vertically, down to $\rm m_{F322W2} \sim 24$: 1G stars are 
$\sim 0.2$ mag bluer than the 2G counterparts, due to the deeper water absorption 
feature centred at $\rm 2.7~\mu m$, which damps the $\rm m_{F322W2}$ flux, thus 
making $\rm (m_{F115W}-m_{F322W2})$ redder. We note a slight trend towards the 
blue in the $\rm 23 < m_{F322W2} < 24$ region, as the effective temperature of 
the stars populating that part of the plane approaches $\rm \sim 2200~K$, where 
the SED peaks in the spectral region at $\rm 1.1-1.2~\mu m$. 

Both the 1G and the 2G main sequences turn to the red for 
$\rm m_{F322W2}$ magnitudes above $\sim 24$, attained by stars
of mass below $\rm 0.08~M_\odot$, characterized by effective
temperatures below $\rm 2000~K$: this is due to the formation 
of condensate species (clouds), which favours the shift of
the SED to longer wavelengths. The 1G and 2G sequences 
return to the blue for $\rm m_{F322W2} > 25.5$, as a consequence
of the gravitational settling of the clouds, which are
gradually removed from the atmosphere.

\begin{figure}
\begin{minipage}{0.98\textwidth}
\resizebox{0.49\hsize}{!}{\includegraphics{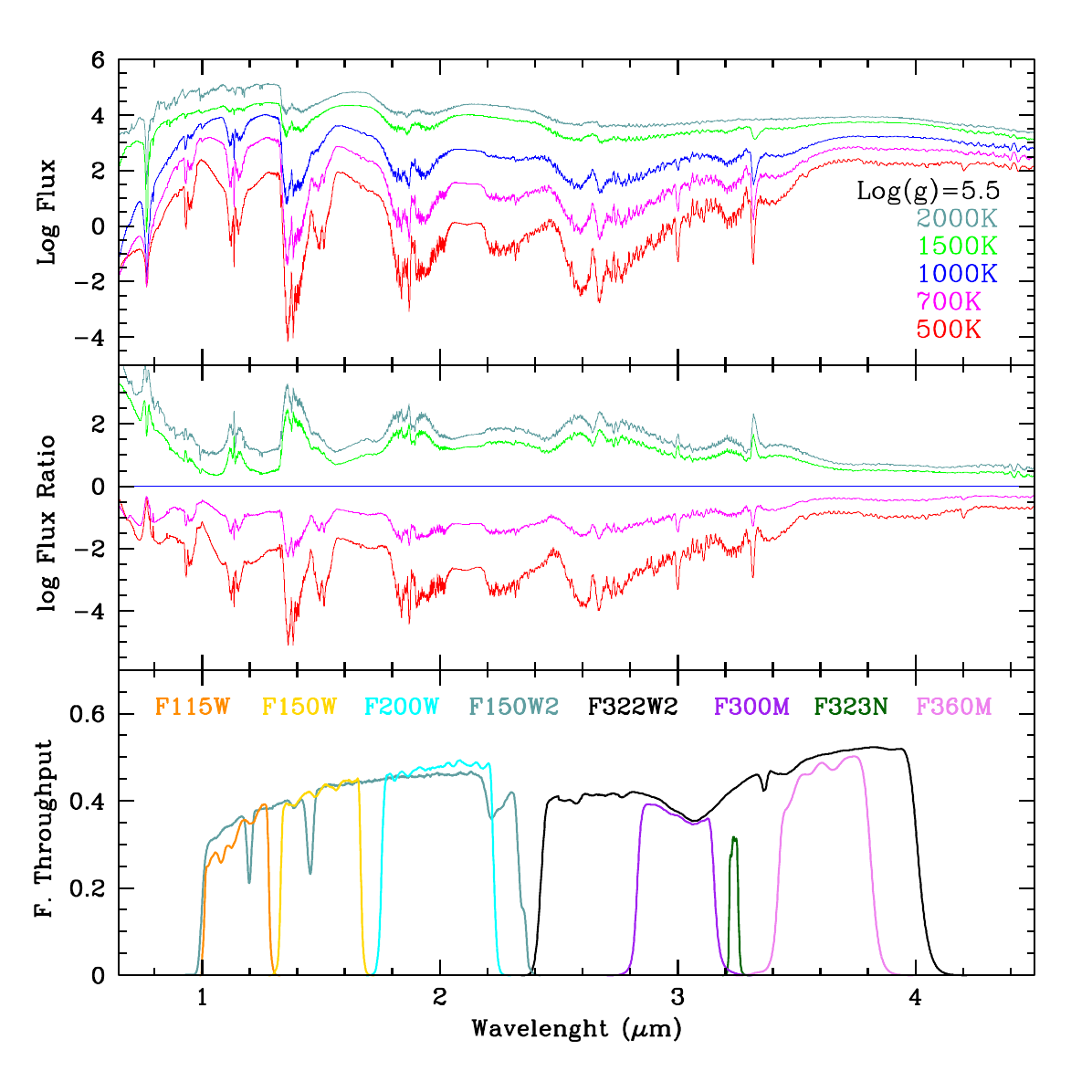}}
\end{minipage}
\caption{\textit{Top panel:} Synthetic spectra of various temperature at the 
labelled gravity; their chemistry and elements distribution correspond to the 
1G case. \textit{Middle panel:} Ratio between the spectra showed in the top 
panel and the one at 1000\,K, taken as reference. \textit{Bottom panel:} 
Throughput of several JWSR filters that are important for the wavelength 
range of our data.}
\label{fspectrapop}
\end{figure}

For a better understanding of the colour excursion of the
isochrones across the $\rm (m_{F115W}-m_{F322W2}, m_{F322W2})$ plane, 
we show in the top panel of Fig.~\ref{fspectrapop} several spectra at 
different temperatures (from 2000\,K to 500\,K and at fixed gravity), 
while in the middle panel we show the logarithm of their ratio (note that 
we have taken the 1000\,K one as reference). We clearly see that the depression 
caused by water molecules gets deeper as the spectra gets colder. To show 
why this is relevant for our data, in the bottom panel we plot the throughputs 
of selected JWST filters\footnote{http://svo2.cab.inta-csic.es/theory/fps/} 
that fall into the wavelength range of the observed photometric catalogues. 
Indeed, as we see from the figure the F322W2 filter intercepts most of the 
depression thus providing a satisfactory explanation to the behaviour we 
see in our models. This behaviour is the 
continuation of the ones already studied in hotter stars (>3000\,K) with 
observed spectra in this cluster \citep{marino24b} and in 
others \citep[see e.g.][ in NGC\,6752 with HST]{milone19}

On general grounds the behaviour of the stellar colours 
M$_{\rm F115W}$-M$_{\rm F322W2}$, 
although not perfectly, is well reproduced by the sets of model atmospheres 
we are using, implying that the way \cite{gera1} include condensate species 
(clouds) in the model is adequate, both to reproduce the shift to the red of 
these colours at $\rm T_{eff}<2000~K$, not predicted by gas--only models, and the 
shift to the blue at $\rm T_{eff}<1500~K$, due to the gravitational settling of 
the clouds, gradually removing them from the atmosphere. Nevertheless, it 
would be too much to infer the goodness of the qualitative difference between 
the two sets of models adopted for the 1G and 2G. In particular, the observed 
main sequence in this plane shows a broadening consistent with the difference 
between the two sets of isochrones, but precisely at $\rm T_{eff}\sim2000~K$ the 1G 
models having $\rm 0.09 \leq M/M_\odot \leq 0.075$ do not seem to have a plausible 
correspondence in the data, as they traverse only a few points of the sample. It looks like the different populations merge into a single sequence corresponding to the 2G models location. This merging appears to occur indeed also in the models, but only after reaching the reddest colours, and the broad data in the magnitude range $\rm M_{F322W}=25-27$\footnote{Although the proper motions of these stars are not available, 
they were considered as cluster members and included in the analysis, 
following \citet{milone23} and MA24, who concluded that the stars in that part of the 
CMD as probable BDs, with some not relevant contamination from the MC, and foreground or 
background objects. Their inclusion is further justified by the identification by MA24 
of two candidate BDs  in that specific range of colours and magnitude values (see their Figure 4). 
Solving this possible ambiguity advocates for further observational explorations by the JWST 
of that region of the CMD in order to obtain accurate proper motion estimates.} seem to correspond to the models shifting to the blue due to the cloud settling.

The left side of Fig.~\ref{doppio} shows the distribution of the stars of
the cluster in the $\rm (m_{F150W}-m_{F322W2}$ , $\rm m_{F322W2})$ plane.
The blue and red lines indicate the 12 Gyr isochrones for the 1G and 2G
stars, respectively, while the full points correspond to the same masses
indicated in Fig.~\ref{fiso}. We note the significant deviation
of the isochrones from the observed sequence in the region of the plane
around $\rm m_{F322W2} \sim 24$, where the isochrones are far from reproducing 
the discontinuity in the observed MS. This discrepancy between the theoretical 
modelling and the data points was recently noticed by
\citet{scalco47tuc2025}, who demonstrated that a satisfactory fit of the
data point in this plane can be obtained only by assuming that the
surface chemical composition of the stars at the age of the cluster changes with 
the stellar mass. This conclusion stems from the fact that reproducing
the observations of the very low MS requires the presence of $\rm CH_4$
molecules, which suggests a smaller surface abundance of oxygen in the
stars of lower mass; this is the sine qua non condition for the formation 
of $\rm CH_4$ molecules at the expenses of CO. \citet{scalco47tuc2025}
proposed that the smaller abundance oxygen might be connected to the
depletion of oxygen atoms onto dust grains. We leave this problem open.

\section{The mass function of 47 Tuc}
\label{massfunction}
\begin{figure}
\centering
\vskip -40pt
\includegraphics[width=0.47\textwidth]{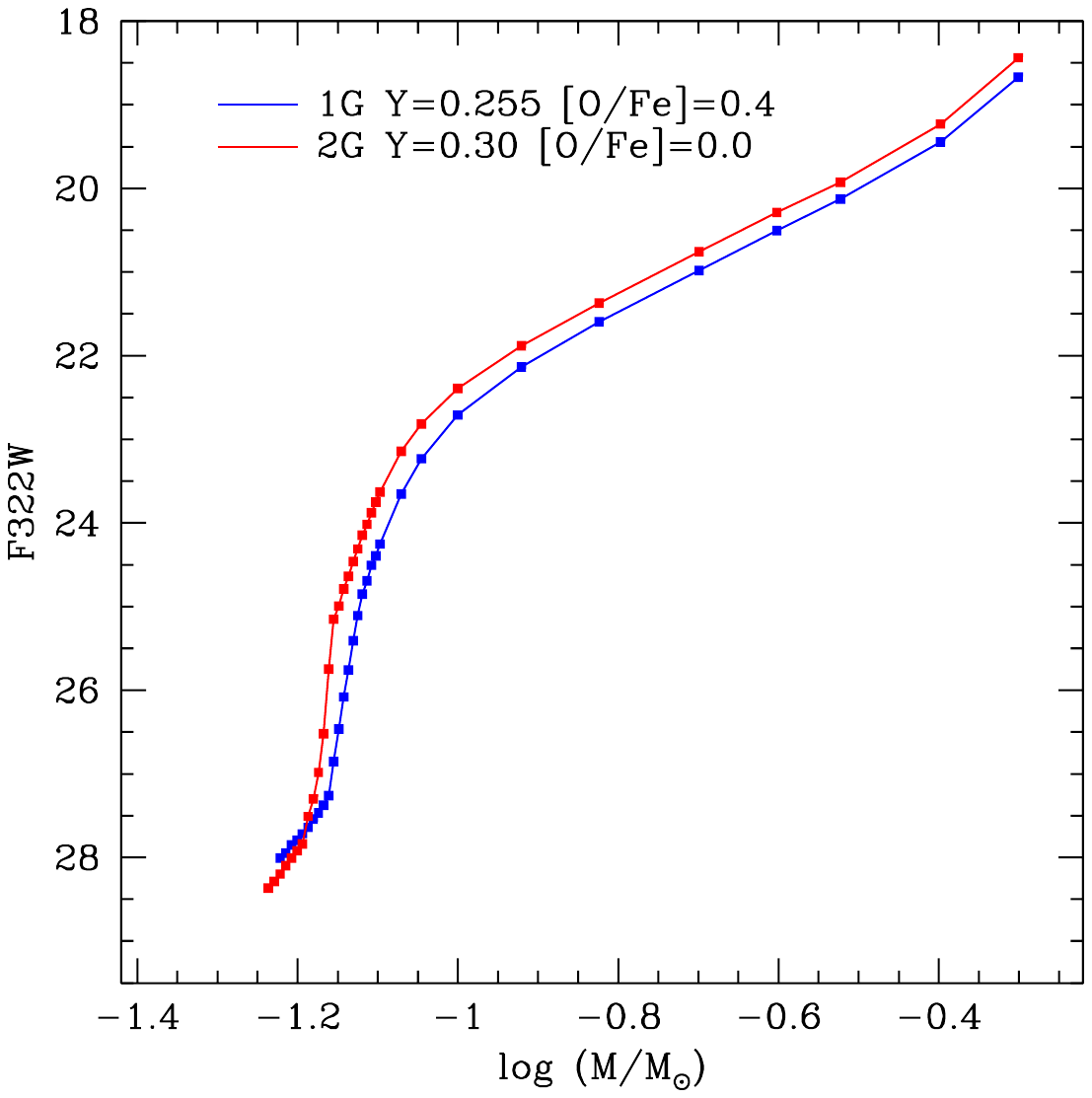}
\vskip -60pt
\caption{ $\rm m_{F322W2}$ versus logarithm of the mass relation defined by
1G (blue line) and 2G (red) stellar models.}
\label{fmlum}
\end{figure}

To build the MF for 47 Tuc, $\rm \Phi(M)$ (where $\rm \Phi(M) \times dM$ gives
the number of stars with mass in the $\rm (M, M+dM)$ range), we convolve the 
luminosity function ($\rm dN/ dm_{F322W2}$, LF) in the $\rm F322W2$ filter by MA24,
based on the number counts of the stars within $\rm m_{F322W2}$ bins 0.4 mag wide, with the 
mass-$\rm m_{F322W2}$ relations obtained for the 1G and
the 2G of the cluster at the age of 12 Gyr, shown in Fig.~\ref{fmlum}. 
Therefore, we have
$$
\rm {\Phi(M) = {dN\over dm_{F322W2}} \times {dm_{F322W2} \over dM}}
$$
The LF considered takes into
account the corrections for completeness, given in MA24.

\begin{figure}
\centering
\vskip -30pt
\includegraphics[width=0.50\textwidth]{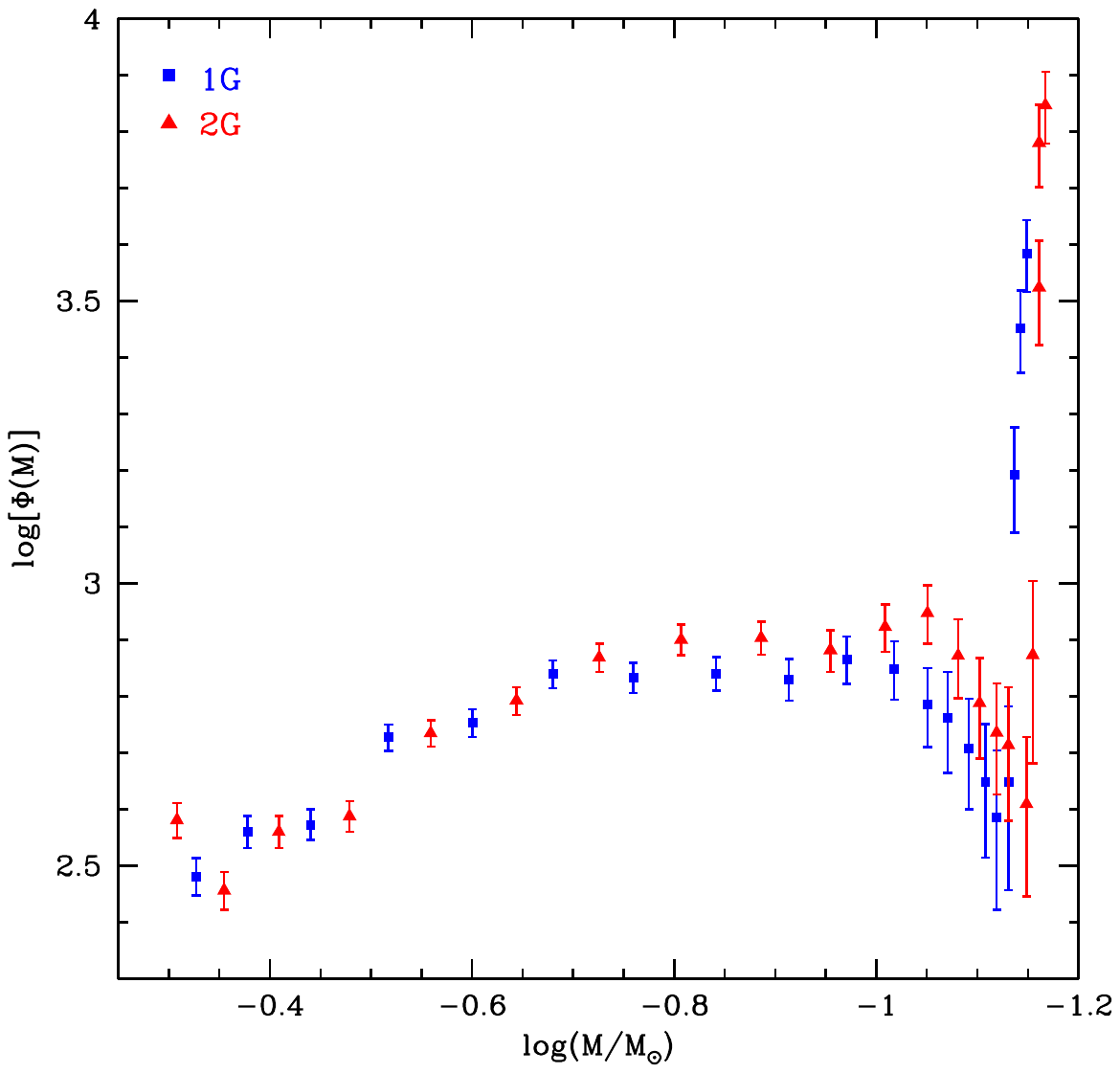}\hfil
\vskip -100pt
\includegraphics[width=0.50\textwidth]{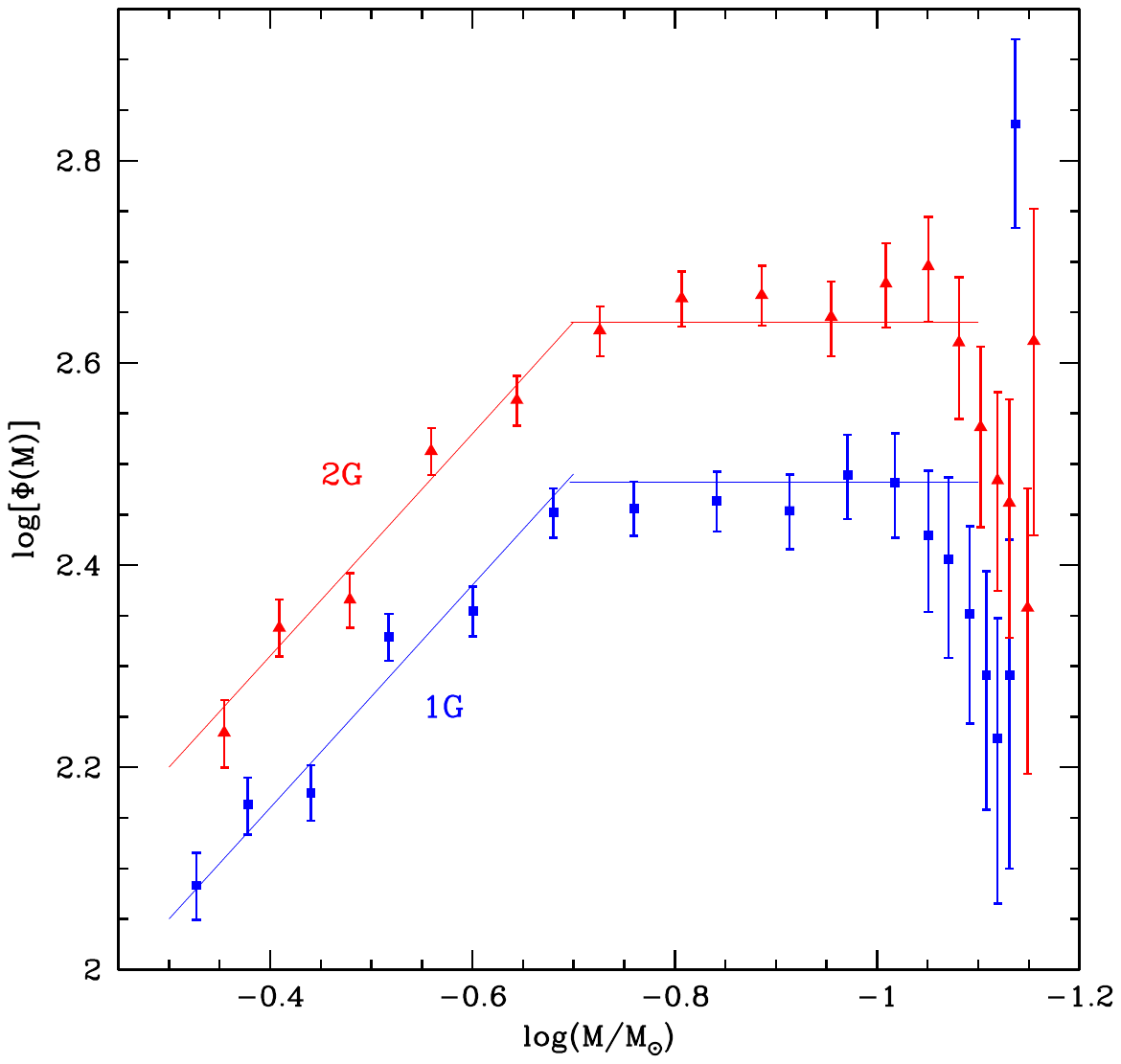}
\vskip-60pt
\caption{Mass function of the two stellar populations of 47 Tuc.
The top panel shows the MF derived under the hypothesis
that all the stars either belong to the 1G (blue dots) or belong to the 2G
(red dots) of the cluster. The bottom panel reports the MF
based on the assumption that $40\%$ of 47 Tuc stars belong to the 1G, 
this percentage increasing up to $46\%$ for $\rm F322W2>21$.
Both MF's are consistent with a Kroupa-like profile with slope
1.1 down to masses $\rm \sim 0.22~M_\odot$, then keep 
constant until the very low-mass domain. 
}
\label{fimf}
\end{figure}

If we assume that all the stars reported in Fig.~\ref{fiso} belong to the
1G or the 2G we obtain the mass function shown in the top panel
of Fig.~\ref{fimf}. For both stellar populations we note an 
increasing trend towards the low-mass domain, down to $\rm \sim 0.2-0.3~M_\odot$,
then an approximately constant MF for $\rm M < 0.2~M_\odot$. The minimum
of the MF at $\rm \sim 0.8~M_\odot$, and the rise in the very low-mass
domain are determined by the oscillating behaviour of the LF, which
exhibits a deep minimum at $\rm m_{F322W2} \sim 25$, then a steep rise
at higher $\rm F322W2$ magnitudes. 

As a further step, we attempted the construction of the separate MF's for the 1G 
and the 2G. Following MA24, we started with the assumption that 1G stars 
account for $42\%$ of the overall population of 47 Tuc, while the
residual $58\%$ is composed of 2G stars. This is averaged between the two fields examined by MA24. 
As a second step, we considered the possibility that these percentages 
can slightly change according to the regions of the plane considered, 
as a possible consequence of the different slopes 
of the mass-$\rm m_{F322W2}$ relations for 1G and 2G stars, which are 
shown in Fig.~\ref{fmlum}: indeed the runs of $\rm m_{F322W2}$ with mass
of the two groups of stars are extremely similar for masses above
$\rm 0.25~M_\odot$, while in the lower mass domain the 
2G relation is steeper, thus in a given $\rm m_{F322W2}$ bin 
the mass range spanned by 2G stars is narrower than the 1G.
Therefore, we allowed the fraction of 1G stars to increase from $42\%$ 
at $\rm m_{F322W2} = 22.5$ to $44\%$ at $\rm m_{F322W2} = 23.5$ (we note here 
that this is still within the errors of the values given by MA24).
In this way, we obtain the MFs shown in the right panel of Fig.~\ref{fimf}. Both 
1G and 2G MFs are consistent with a Kroupa-like profile slope 
1.1.

Regarding the $\rm M < 0.08~M_\odot$ mass range, both in the 
1G and 2G cases, we note a significant increase in the MF, 
which reflects the notable increase in the number of sources populating 
the faintest $\rm m_{F322W2}$ bins, as is clear in Fig.~\ref{fiso}.

The last points of the mass functions shown in Fig.~\ref{fimf} rely on the hypothesis that the increase in number of stars shown by the empirical LF is indeed real as found by MA24. Obviously it is very awkward to believe in a sharp increase of the mass function in a very small range of brown dwarf masses, unless they correspond to the transition region from stars to brown dwarfs, as suggested by MA24, where the mass luminosity relation has a sharp change of slope. 
Let us then assume that the mass function is flat for $\rm M<0.2~M_\odot$, as shown in the bottom panel of Fig.~\ref{fimf}, and look at the derivative of the mass--F322W2 magnitude relation of the models. The result, for both 1G and 2G models, is shown on the right side of Fig.~\ref{doppio}, normalized to the upper part of the luminosity function. We see that the behaviour of stellar models imply a sharp decrease in the number of stars during the transition from the stellar to the brown dwarf regime, but, at smaller masses, the brown dwarf simple cooling causes a new increase of the LF, similar to the increase of the empirical LF, but occurring $\sim$\,1.5\,mag dimmer. In view of the uncertainties in the models, we suggest that it is  possible that the transition from the stellar to the BD regime is indeed what we are witnessing in the data. If the small mass interval of the transition masses is indeed located at m$_{F322W} \sim 24$, it could also better justify the apparent lack of 1G stars noticed when discussing Fig.\,\ref{fiso}.  

The present findings are based on empirical arguments related to the continuity 
of the MF of 1G and 2G populations of 47 Tuc. For a more reliable calculation of the
MF in the very low-mass domain, both a theoretical effort in the study of model atmospheres 
and a further observational effort for membership determination by proper motions of the
faintest stars populating the bottom of the main sequence are needed.


\section{Conclusions}
\label{concl}

We used recent JWST data of 47 Tuc to study the bottom of the main sequence of the cluster, 
with particular care to the still poorly explored stellar/sub-stellar transition.
To this aim, we calculated evolutionary sequences of $\rm M\geq 0.06~M_\odot$ stars
with metallicity $\rm [Fe/H]=-0.78$, based on the non-grey
treatment of the atmospheric and sub-atmospheric layers. The different
stellar populations of 47 Tuc were investigated by adopting 
oxygen with $\rm [O/Fe]=+0.4$, to represent the 1G, and 
stellar models with $\rm [O/Fe]=0$, to simulate the 2G
of the cluster. 

We discuss in detail the transition between stars ---masses fully substained by p-p burning--- and brown dwarfs ---masses which do not ignite hydrogen due to electron degeneracy. The intermediate transition masses are supported by p-p burning for billions of years, but finally cool as brown dwarfs, as shown in seminal computations by \citet{dm1985}. We show that the mass range covered by transition masses is very small, and this produces a dip in the derivative of the mass luminosity relation.

We find a slight mass difference between the threshold value separating the stellar 
and the sub-stellar domains, the last ones permanently burning central hydrogen 
are $\rm 0.074~M_\odot$ for the 1G and $\rm 0.071~M_\odot$ for the 2G. 
A satisfactory agreement between the isochrones and the data points in the 
$\rm (m_{F115W}-m_{F322W2}$, $\rm m_{F322W2})$ colour-magnitude diagram is obtained 
with reddening $\rm E(B-V)=0.03$ and distance modulus $\rm (m-M)_0=13.21$.
Theoretical isochrones successfully reproduce colour spread of the MS 
in the $\rm m_{F322W2}<23$ region of the plane, as well as the upturn to the
red in the fainter region of the diagram. In particular, these results
indicate that the $\rm 1.3 < (m_{F115W}-m_{F322W2}) < 2.2$ region at
$\rm m_{F322W2} > 24$ is populated by $\rm M \leq 0.08~M_\odot$ stars, 
which represent the gradual transition from the stellar to sub-stellar domain.
We speculate on the possibility that this transition from the star to brown dwarf domain actually occurs at larger luminosity: this would provide better agreement between the observed colour magnitude diagrams and the isochrones and a better agreement with the rising of the LF at the dimmest magnitudes, and with the apparent disappearance of the 1G sequence when the MS rapidly turns to redder colours.

\begin{acknowledgements}
PV acknowledges support by the INAF-Theory-GRANT 2022 
“Understanding mass loss and dust production from evolved stars”.\\
CV acknowledges support by the INAF-fellowship 2024 
“Asteroseismological properties of red giant and clump stars” and 
support by the INAF-fellowship 2025 “Web services and imaging data reduction for LBT”.
\end{acknowledgements}

%
%
\bibliographystyle{aa}
\bibliography{browndwarfs}

\begin{appendix}
\section{}
Table A.1 and Table A.2 below show the values for 47 Tuc 1G and 2G isochrones, based on stellar evolutionary models of different masses (specifications in their captions, respectively).

\begin{table*}[htb]
  \centering
  \setlength{\tabcolsep}{10pt}
  \renewcommand{\arraystretch}{1.5}
  \begin{tabular}{c c c c c c c c}
\hline 
$\rm M/M_\odot$ & $\rm T_{eff}$ & $\rm log(L/L_\odot)$ & $\rm m_{F322W2}$ & $\rm m_{F115W}-m_{F322W2}$ & $\rm m_{F150W2}-m_{F322W2}$ & $\rm ({L_{pp}/L})_{12 Gyr}$ & $\rm ({L_{pp}/L})_{max}$ \\
\hline
$0.060$ & $858.0$ & $-5.56$ & $14.70$ & $1.90$ & $1.95$ & $0.0014$ & $0.08$ \\
$0.061$ & $878.5$ & $-5.52$ & $14.64$ & $1.85$ & $1.90$ & $0.0019$ & $0.10$ \\
$0.062$ & $899.0$ & $-5.48$ & $14.54$ & $1.82$ & $1.82$ & $0.0025$ & $0.12$ \\
$0.063$ & $923.5$ & $-5.44$ & $14.49$ & $1.80$ & $1.80$ & $0.0032$ & $0.14$ \\
$0.064$ & $947.0$ & $-5.39$ & $14.41$ & $1.75$ & $1.72$ & $0.0043$ & $0.17$ \\
$0.065$ & $970.5$ & $-5.35$ & $14.33$ & $1.73$ & $1.68$ & $0.0057$ & $0.21$ \\
$0.066$ & $996.5$ & $-5.31$ & $14.23$ & $1.70$ & $1.63$ & $0.008$ & $0.25$ \\
$0.067$ & $1022.9$ & $-5.27$ & $14.16$ & $1.65$ & $1.63$ & $0.012$ & $0.30$ \\
$0.068$ & $1047.7$ & $-5.22$ & $14.06$ & $1.62$ & $1.63$ & $0.019$ & $0.36$ \\
$0.069$ & $1079.3$ & $-5.07$ & $13.95$ & $1.60$ & $1.62$ & $0.034$ & $0.41$ \\
$0.070$ & $1206.4$ & $-4.97$ & $13.54$ & $1.91$ & $1.61$ & $0.047$ & $0.50$ \\
$0.071$ & $1292.7$ & $-4.84$ & $13.16$ & $2.00$ & $1.55$ & $0.11$ & $0.60$ \\
$0.072$ & $1359.2$ & $-4.73$ & $12.77$ & $2.31$ & $1.50$ & $0.31$ & $0.70$ \\
$0.073$ & $1447.4$ & $-4.58$ & $12.45$ & $2.37$ & $1.46$ & $0.72$ & $0.82$ \\
$0.074$ & $1580.9$ & $-4.39$ & $12.09$ & $1.95$ & $1.25$ & $0.95$ & $0.97$ \\
$0.075$ & $1719.5$ & $-4.22$ & $11.80$ & $1.57$ & $1.10$ & $1.00$ & $1.00$ \\
$0.076$ & $1828.5$ & $-4.096$ & $11.55$ & $1.41$ & $0.95$ & $1.00$ & $1.00$ \\
$0.077$ & $1937.7$ & $-3.97$ & $11.38$ & $1.25$ & $0.81$ & $1.00$ & $1.00$ \\
$0.078$ & $2024.9$ & $-3.88$ & $11.20$ & $1.15$ & $0.75$ & $1.00$ & $1.00$ \\
$0.079$ & $2100.4$ & $-3.80$ & $11.08$ & $1.06$ & $0.69$ & $1.00$ & $1.00$ \\
$0.080$ & $2172.2$ & $-3.73$ & $10.94$ & $1.08$ & $0.66$ & $1.00$ & $1.00$ \\
$0.085$ & $2457.0$ & $-3.46$ & $10.34$ & $1.04$ & $0.60$ & $1.00$ & $1.00$ \\
$0.090$ & $2648.5$ & $-3.28$ & $9.92$ & $1.07$ & $0.59$ & $1.00$ & $1.00$ \\
$0.100$ & $2875.4$ & $-3.04$ & $9.40$ & $1.09$ & $0.58$ & $1.00$ & $1.00$ \\
$0.120$ & $3069.0$ & $-2.79$ & $8.82$ & $1.10$ & $0.57$ & $1.00$ & $1.00$ \\
$0.150$ & $3202.6$ & $-2.55$ & $8.29$ & $1.09$ & $0.56$ & $1.00$ & $1.00$ \\
$0.200$ & $3321.2$ & $-2.28$ & $7.67$ & $1.07$ & $0.54$ & $1.00$ & $1.00$ \\
$0.250$ & $3410.4$ & $-2.08$ & $7.20$ & $1.06$ & $0.52$ & $1.00$ & $1.00$ \\
$0.300$ & $3485.8$ & $-1.91$ & $6.82$ & $1.06$ & $0.51$ & $1.00$ & $1.00$ \\
$0.400$ & $3652.6$ & $-1.62$ & $6.14$ & $1.08$ & $0.50$ & $1.00$ & $1.00$ \\
$0.500$ & $3931.0$ & $-1.28$ & $5.36$ & $1.10$ & $0.49$ & $1.00$ & $1.00$ \\
\hline
\end{tabular}
\vskip10pt
\caption{The table lists various theoretical and observational properties of stellar models
of different mass (reported in col. 1) with the chemical composition used to
model the 1G of the cluster, namely $\rm [Fe/H]=-0.78$, helium mass fraction
$\rm Y=0.255$, and $\rm [O/Fe]=+0.4$. Col. 2 and 3 give the effective temperature and
luminosity, whereas col. from 4 to 6 report $\rm m_{F322W2}$ magnitude, the $\rm (m_{F115W}-m_{F322W2})$ and 
$\rm (m_{F150W}-m_{F322W2})$ colours, respectively. Col. 7 and 8 give the fractional p-p
luminosity at the age of 12 Gyr and the largest p-p
luminosity reached during the evolution.
}
\label{tab:teflum1g}
\end{table*}

\begin{table*}[htb]
  \centering
  \setlength{\tabcolsep}{10pt}
  \renewcommand{\arraystretch}{1.5}
  \begin{tabular}{c c c c c c c c}
\hline 
$\rm M/M_\odot$ & $\rm T_{eff}$ & $\rm log(L/L_\odot)$ & $\rm m_{F322W2}$ & $\rm m_{F115W}-m_{F322W2}$ & $\rm m_{F150W2}-m_{F322W2}$ & $\rm ({L_{pp}/L})_{12 Gyr}$ & $\rm ({L_{pp}/L})_{max}$ \\
\hline
$0.060$ & $874.0$  & $-5.57$ & $14.89$ & $2.01$ & $1.95$ & $0.0015$ & $0.15$ \\
$0.061$ & $899.5$  & $-5.52$ & $14.79$ & $1.93$ & $1.86$ & $0.0019$ & $0.18$ \\
$0.062$ & $926.0$  & $-5.47$ & $14.70$ & $1.85$ & $1.80$ & $0.0029$ & $0.22$ \\
$0.063$ & $954.5$  & $-5.42$ & $14.61$ & $1.79$ & $1.74$ & $0.0038$ & $0.27$ \\
$0.064$ & $979.5$  & $-5.38$ & $14.53$ & $1.75$ & $1.71$ & $0.0054$ & $0.32$ \\
$0.065$ & $1038.2$ & $-5.26$ & $14.20$ & $1.70$ & $1.70$ & $0.0082$ & $0.42$ \\
$0.066$ & $1086.0$ & $-5.18$ & $13.99$ & $1.83$ & $1.70$ & $0.014$ & $0.50$ \\
$0.067$ & $1161.2$ & $-5.06$ & $13.67$ & $1.94$ & $1.70$ & $0.026$ & $0.60$ \\
$0.068$ & $1272.6$ & $-4.89$ & $13.21$ & $2.16$ & $1.68$ & $0.063$ & $0.72$ \\
$0.069$ & $1472.6$ & $-4.58$ & $12.44$ & $2.40$ & $1.52$ & $0.25$  & $0.85$ \\
$0.070$ & $1607.3$ & $-4.37$ & $11.85$ & $2.34$ & $1.42$ & $0.93$  & $0.90$ \\
$0.071$ & $1712.2$ & $-4.29$ & $11.79$ & $2.11$ & $1.34$ & $1.00$  & $1.00$ \\
$0.072$ & $1809.6$ & $-4.15$ & $11.48$ & $1.79$ & $1.19$ & $1.00$  & $1.00$ \\
$0.073$ & $1917.2$ & $-4.03$ & $11.33$ & $1.57$ & $1.03$ & $1.00$  & $1.00$ \\
$0.074$ & $2011.4$ & $-3.93$ & $11.15$ & $1.47$ & $0.94$ & $1.00$  & $1.00$ \\
$0.075$ & $2099$   & $-3.85$ & $11.00$ & $1.38$ & $0.86$ & $1.00$  & $1.00$ \\
$0.076$ & $2187.8$ & $-3.76$ & $10.84$ & $1.33$ & $0.80$ & $1.00$  & $1.00$ \\
$0.077$ & $2269.3$ & $-3.69$ & $10.71$ & $1.29$ & $0.75$ & $1.00$  & $1.00$ \\
$0.078$ & $2349.6$ & $-3.62$ & $10.57$ & $1.28$ & $0.72$ & $1.00$  & $1.00$ \\
$0.079$ & $2415.5$ & $-3.56$ & $10.44$ & $1.27$ & $0.70$ & $1.00$  & $1.00$ \\
$0.080$ & $2482.0$ & $-3.50$ & $10.32$ & $1.27$ & $0.69$ & $1.00$  & $1.00$ \\
$0.085$ & $2723.3$ & $-3.28$ & $9.84$  & $1.29$ & $0.66$ & $1.00$  & $1.00$ \\
$0.090$ & $2884.0$ & $-3.13$ & $9.51$  & $1.29$ & $0.64$ & $1.00$  & $1.00$ \\
$0.100$ & $3066.2$ & $-2.93$ & $9.08$  & $1.28$ & $0.63$ & $1.00$  & $1.00$ \\
$0.120$ & $3221.1$ & $-2.70$ & $8.57$  & $1.27$ & $0.62$ & $1.00$  & $1.00$ \\
$0.150$ & $3338.9$ & $-2.48$ & $8.06$  & $1.25$ & $0.60$ & $1.00$  & $1.00$ \\
$0.200$ & $3449.1$ & $-2.21$ & $7.45$  & $1.23$ & $0.59$ & $1.00$  & $1.00$ \\
$0.250$ & $3536.7$ & $-2.01$ & $6.98$  & $1.22$ & $0.58$ & $1.00$  & $1.00$ \\
$0.300$ & $3608.3$ & $-1.85$ & $6.62$  & $1.23$ & $0.57$ & $1.00$  & $1.00$ \\
$0.400$ & $3789.7$ & $-1.54$ & $5.92$  & $1.20$ & $0.55$ & $1.00$  & $1.00$ \\
$0.500$ & $4127.6$ & $-1.17$ & $5.13$  & $1.14$ & $0.48$ & $1.00$  & $1.00$ \\
\hline
\end{tabular}
\vskip10pt
\caption{The table lists the same properties reported in Table~\ref{tab:teflum1g}, for the stellar
models calculated with the chemical composition assumed for the 2G of 47 Tuc, namely
$\rm [Fe/H]=-0.78$, helium mass fraction $\rm Y=0.30$, and $\rm [O/Fe]=0$.
}
\label{tab:teflum2g}
\end{table*}

\end{appendix}

\end{document}